 \documentstyle[preprint,aps]{revtex}
 \begin{document}
 \draft
 \baselineskip 24 true pt
 \vsize=9.5 true in \voffset=.2 true in
 \hsize=6.0 true in \hoffset=.25 true in

 \newpage
 \begin{center}
 {\Large{\bf Jahn-Teller Coupled Charge Density Wave in A Two Orbital
 Double
 Exchange System}}
 \end{center}
 \vskip 1.0cm
 \begin{center}
 \end{center}
 \vskip 0.50cm
 \begin{center}
 {Manidipa Mitra\footnote{ e-mail: mmitra@bose.res.in}}
 \end{center}

 \begin{center}
 {\em S. N. Bose National Centre for Basic Sciences, JD Block, \\
 Sector III, Salt Lake City, Kolkata - 700 098, India}\\

 \end{center}

 \vskip 1.0cm

 PACS No. 63.20.-e,  71.45.Lr,  71.20.Be  
 \vskip 1.0cm
 \begin{center}
 {\bf Abstract}
 \end{center}
 \vskip 0.5cm

 In a two orbital double exchange model we formulate Jahn-Teller coupled
 charge density wave in one electron per lattice site limit. Softening of 
 Jahn-Teller phonons corresponding to distortion modes $Q_2$ or $Q_3$ 
 associated with perfect nesting of Fermi
 surface leads to this instability at low temperature.
 The gap equation for charge density wave state and its
 dependences on electron-lattice coupling are calculated explicitly
 when any one of the Jahn-Teller modes is excited cooperatively.
 We find that the $Q_2$ distortion mode yields
 lowest free energy. The calculated heat capacity for each mode shows
 $T^3$ behavior at low temperature. Effect of
 electron-lattice interaction on collective mode, such as
 amplitude mode, is more pronounced when the excited mode is $Q_2$.

\vskip 1. cm

\newpage
\section{Introduction}
 It is widely accepted that the collective nature of coupling of
 $e_g$ electrons to the underlying lattice degrees of freedom
 have great impact on extraordinary
 properties of the colossal magneto resistive (CMR) manganites
 \cite{toku}. The parent
 compound $LaMnO_3$ contains Jahn-Teller (JT) ions $Mn^{3+}$, which is
 orbitally
 degenerate. The on-site vibrational modes of the $MnO_6$ octahedra
 are breathing mode $Q_1$ and the Jahn-Teller modes -- $Q_2$
 (in-plane distortion mode), $Q_3$ (octahedral stretching mode).
 Due to the coupling between the $e_g$ electrons and
 $Q_2$, $Q_3$  modes the degeneracy of $Mn^{3+}$ is lifted. In the
 lattice the $MnO_6$ octahedra are co-operatively stretched out, resulting
 a C - type orbital and A - type spin antiferromagnet structure
 \cite{mura}.
 Several recent experimental investigations such as
 high-resolution electron microscopy \cite{naga}, THz time-domain
 spectroscopy
 \cite{nori}, high-resolution ARPES \cite{chuang}, non-linear electrical
 response
 \cite{wahl}, x-ray and neutron scattering \cite{arg,camp} measurements
 reveal the evidence
 of charge density waves in different composition of CMR manganites.
 Moreover, ARPES results \cite{chuang} show Fermi surface nesting, which
 ensures
 the possibility of density wave formation in these materials.
 It is well known that in a low dimensional one orbital system,
 due to particular geometry of the Fermi surface,
 electron-phonon interactions lead to periodic lattice distortions. This
 results a ground state characterized by a gap in the single particle
 excitation spectrum at low temperature, which is known as charge density
 wave (CDW) \cite{grun}.
 In CMR manganites the Jahn-Teller vibrational modes play major role to
 shape the properties of these systems compared to the breathing mode
 phonon
 $Q_1$ which does not alter the symmetry of the $MnO_6$ octahedron.
 The distribution of charge density
 at two $e_g$ orbitals, $d_{x^2-y^2}$ and $d_{3z^2-r^2}$, depends on
 excited JT mode.
 In presence of these two orbitals, the electron can be either
 in any one of the orbitals or in any linear combination of two
 \cite{pereb}.
 It appears that the
 perfect nesting of the Fermi surface associated with Jahn-Teller
 distortions
 may give rise to possibilities of occurrence of charge density waves in
 these systems. Therefore, the resulting
 charge density wave will be coupled to the orbital order of the
 system. We denote this density wave as Jahn-Teller coupled charge density
 wave (JTCCDW).

 In the present work, we investigate the formation of Jahn-Teller coupled
 charge density
 wave instability in two orbital, double exchange model with one electron
 per site. The paper is organized as follows.
 In section II we formulate the Hamiltonian for Jahn-Teller coupled
 charge density wave in two dimension and
 calculate the JTCCDW order parameter for different JT vibrational modes.
 To determine the dominant JT distortion mode,
 we  calculate the free energy and hence heat capacity in the
 JTCCDW state.  In the same spirit of
 CDW system, the spectral density functions of the ordered ground state
 are calculated from the amplitude mode response function of JTCCDW state
 generated due to each JT active mode. The results of our
 calculations are presented in section III. Lastly, in section IV we make some
 concluding remarks.

\section{ Formalism }
 We consider two orbital double exchange model in two dimension, and
 large Hund's rule coupling limit in which the ground state configuration
 corresponds to the itinerant spin being parallel to the core spin at each
 site.  If one compares our model
 with the manganite (e.g. $LaMnO_3$) system the core spins and the
 itinerant electrons may be identified with the $t_{2g}^3$ (localized)
 electrons and the $e_{g}^1$ (mobile) electrons of $Mn^{3+}$ ion
 respectively. The kinetic part of the Hamiltonian in momentum representation 
 is given by

 \begin{eqnarray}
 H_1 = \sum_k B_k^\dagger {\bf T} B_k,
 \label{eqn1}
 \end{eqnarray}
 where $B_k^\dagger \equiv ( d_{1,k}^\dagger,~ d_{2,k}^\dagger)$ with
 $d_{1}$ and $d_{2}$ being
 the $e_g$ - electron annihilation operators in two orthonormal orbitals
 $d_{x^2 - y^2}$ and $d_{3z^2 - r^2}$ respectively.
 The elements of the hopping matrix ${\bf T}$ are given by
 ${\bf T}_{1,1}=-1.5 {\tilde t}[\cos k_x + \cos k_y]$,
 ${\bf T}_{2,2}=-0.5 {\tilde t} [\cos k_x + \cos k_y]$, and
 ${\bf T}_{1,2}=0.5 \sqrt{3} {\tilde t} [\cos k_x - \cos k_y ]$.
 The transfer hopping integral $t$ is modified as $\tilde t = t ~ \cos
 (\frac{\theta}{2})$ because of double exchange interaction \cite{and}. 
 If the $t_{2g}$ spins are assumed classical then $\theta $ is the relative
angle between two neighbouring spins. Thus, in case of ferromagnetic order (FM)
$\tilde t = t$. We consider that the $A$-type spin antiferro order has set in.
Due to strong Hund's rule coupling the transport is restricted to spin polarizedelectrons in two dimensions only. So the spin degrees of freedom may be omitted in the Hamiltonian without loss of generality.

 The Jahn-Teller coupling parts of  the Hamiltonian for $Q_2$ and $Q_3$
 mode can be written as \cite{pereb}
 \begin{eqnarray}
 H_{Q_2} = g \sum _i Q_{2i} (d_{1i}^\dagger d_{2i} + d_{2i}^\dagger
 d_{1i}),
 \label{eqn2a}
 \end{eqnarray}
 \begin{eqnarray}
 H_{Q_3} = g \sum _i Q_{3i} (d_{1i}^\dagger d_{1i} - d_{2i}^\dagger
 d_{2i}),
 \label{eqn2b}
 \end{eqnarray}
 where $g$ is the electron-JT phonon coupling strength.
 To diagonalize the kinetic part of the Hamiltonian appearing in
 equation (\ref {eqn1}) we introduce the new fermionic operators
 $c^1_{k}$ and $c^2_{k}$ as

 \begin{eqnarray}
 c^1_{k} &=& \sin \frac{\theta_k}{2}~ d_{1,k} + \cos \frac{\theta_k}{2}~
 d_{2,k}  ,\nonumber\\
 c^2_{k} &=&
 \cos \frac{\theta_k}{2}~ d_{1,k} - \sin \frac{\theta_k}{2}~ d_{2,k},
 \label{eqn3}
 \end{eqnarray}
 with  $\tan \theta_k = \frac {{\sqrt 3} (\cos ~ k_x - \cos ~ k_y)}
 {(\cos~ k_x + \cos~ k_y)}$.
 The eigen values of the kinetic energy are given by
 $\lambda_{n,k} = -\cos k_x - \cos k_y - (-1)^n \sqrt{\cos^2 k_x
 + \cos^2 k_y  - \cos k_x  \cos k_y }$ where $n=1,2$.
 The Fermi sea corresponding to $\lambda_{1,k}$ is
 the intersection of the region
 $-\pi/2 \le  k_x \le \pi/2$ with the region
 $-\pi/2 \le  k_y \le \pi/2$ and corresponding to
 $\lambda_{2,k}$ is the union of the region
 $-\pi/2 \le  k_x \le \pi/2$ (with all values of $k_y$ allowed) to
 the region $-\pi/2 \le  k_y \le \pi/2$ (with all values of $k_x$ allowed).
 It is evident from the symmetry of the bands that at half-filling
 $\lambda_{1(2),\vec k}~ = - ~ \lambda_{2(1),(\vec k + \vec Q )}$
 for $\vec Q = (\pi, \pi)$, which shows a perfect nesting at wave vector
 $\vec Q$.

 The lattice instability in low dimensional system is usually driven
 by the softening of the wave vector $\vec Q$ due to electron-phonon
 (e-ph) interactions, which is called Kohn anomaly. This is a precursor to
 the Peierls transition. Peierls instability brings about the formation
 of the CDW with an accompanying periodic lattice distortion which
 in turn gives rise to a band gap at the Fermi level driving the
 system from a conductor to a insulator. In order to find
 how the JT phonon $Q_2$ or $Q_3$ softens to set similar instability,
 it is needed to calculate the phonon self-energy arising from
 e-ph interaction for $Q_2$ or $Q_3$ mode. The phonon response
 function is determined by the phonon Green's function given
 by \cite{mahan},

 \begin{eqnarray}
 D_{q,q^\prime} (q, q^\prime, \omega) =
 \frac{\omega_q~ \delta_{-q,q^\prime}}{\pi \left[\omega^2 - \omega_q^2 -
 \omega_q \Pi (q, q^\prime, \omega)\right]}.
 \label{eqn4}
 \end{eqnarray}
 Here $\Pi(q,\omega)$ is the self-energy acquired by the $q$ -th phonon
 and $\omega_q$ is the bare phonon frequency. The phonon self-energy
 is given by
 \begin{eqnarray}
 \Pi(q, \omega) =  4 \pi^2 g^2 \chi ( q, \omega),
 \label{eqn5}
 \end{eqnarray}
 where $\chi ( q, \omega)$ is the density response function.
 We denote $\chi (q, \omega)$ as $\chi_2 (q, \omega)$
 and $\chi_3 (q, \omega)$ respectively for $Q_2$ and $Q_3$ distorted modes.

 The phonon response function for $Q_2$ and $Q_3$ mode can be written as
 \cite{sudha},

 \begin{eqnarray}
 \chi_2 (q, \omega) &=& \sum _k~ \langle \langle~ \rho_2(q); \rho_2(-q)~
 \rangle \rangle _\omega,
 \nonumber\\
 \chi_3 ( q, \omega) &=&  \sum _k~ \langle \langle~ \rho_3(q); \rho_3(-q)
 ~\rangle \rangle~_\omega
 \label{eqn6}
 \end{eqnarray}
 where $\rho_2(q)=(d_{1,k+q}^\dagger d_{2k} + d_{2,k+q}^\dagger d_{1k})$
 and  $\rho_3(q)=(d_{1,k+q}^\dagger d_{1k} - d_{2,k+q}^\dagger d_{2k})$.
 The renormalized phonon frequency with the wave vector $q=\mid \vec Q
 \mid$
 corresponding to the $Q_2$ or $Q_3$ mode can be determined from
 the pole of the phonon Green's function,
 with the static response functions $\chi_2(\vec Q, 0)$ or $\chi_3(\vec
 Q,0)$
 respectively. It can be shown \cite{mm} that
 $\chi_2(\vec Q, 0)$ and $\chi_3(\vec Q, 0)$ diverges at $T \rightarrow 0$.

 The response function $\chi_2 (\vec Q, 0)$ diverges faster than
 $\chi_3 (\vec Q,0)$ \cite{mm}. The divergence of the response functions
 with decreasing temperature indicate the softening of $Q_2$
 and $Q_3$ vibrational modes and hence a possibility of Fermi surface
 instability. The temperature at which the soft mode frequency for $Q_2$ or 
 $Q_3$ vanishes is the Jahn-Teller coupled charge density wave transition 
 temperature $T_{CDW}$ corresponding to that mode.

 In this report, we formulate the Jahn-Teller coupled charge density wave
 when either $Q_2$ or $Q_3$ mode gets excited cooperatively.
 For this purpose we use (\ref {eqn3}) and second quantized form
 of the phonon modes given by
 $(b^{2(3)}_{\vec q} + {b^{2(3)}_{-\vec q}}^\dagger)$
 for the wave vector $\vec q = \vec Q$ in equations (\ref {eqn2a}) and
 (\ref {eqn2b}). By using canonical transformation \cite{kittel}, we
 derive the effective electron-electron interaction mediated by
 e-ph interaction as,
 \begin{eqnarray}
 H^{e-e}_{Q_2} &=& -  \sum _{kk^\prime} \mid \tilde g_{kk^\prime} \mid
 \sin \theta_{k^\prime} ~
 ( c_{k^\prime+Q}^{1\dagger} c_{k^\prime} ^2 + c_{k^\prime}^{2\dagger}
 c_{k^\prime+Q} ^1 +
 c_{k^\prime}^{1\dagger} c_{k^\prime+Q} ^2 +  c_{k^\prime+Q}^{2\dagger}
 c_{k^\prime} ^1)
 \nonumber \\
 &\times & \sin \theta_k ~ ( c_{k+Q}^{1\dagger} c_k ^2 +
 c_{k}^{2\dagger} c_{k+Q} ^1 +
 c_{k}^{1\dagger} c_{k+Q} ^2 +  c_{k+Q}^{2\dagger} c_{k} ^1),
 \label{eqn7a}
 \end{eqnarray}
 \begin{eqnarray}
 H^{e-e}_{Q_3} &=& -  \sum _{kk^\prime} \mid \tilde g_{kk^\prime}
 \mid \cos \theta_{k^\prime}  ~
 ( c_{k^\prime+Q}^{1\dagger} c_{k^\prime} ^2 + c_{k^\prime}^{2\dagger}
 c_{k^\prime+Q} ^1 +
 c_{k^\prime}^{1\dagger} c_{k^\prime+Q} ^2 +  c_{k^\prime+Q}^{2\dagger}
 c_{k^\prime} ^1)
 \nonumber \\
 &\times & \cos \theta_k ~( c_{k+Q}^{1\dagger} c_k ^2 +
 c_{k}^{2\dagger} c_{k+Q} ^1 +
 c_{k}^{1\dagger} c_{k+Q} ^2 +  c_{k+Q}^{2\dagger} c_{k} ^1),
 \label{eqn7b}
 \end{eqnarray}
 where $\tilde g_{kk^\prime}$ is considered as average interaction strength
 and $\tilde g_{kk^\prime}=-\mid \tilde g_{kk^\prime} \mid$.
 It is to be mentioned here, while deriving the equations (\ref {eqn7a} ),
 (\ref {eqn7b}) we have considered only terms like
 $c^{1(2)\dagger}c^{2(1)}$
 in the electron-phonon interaction.

 Now we write the total Hamiltonian corresponding to $Q_2$ and $Q_3$ as,
 \begin{eqnarray}
 H^{Q_2(Q_3)} = H_1 + H^{Q_2(Q_3)}_{e-e}.
 \label{eqntotal}
 \end{eqnarray}
 In the mean-field approximation,
 the JTCCDW order parameter ($\Delta_k$) for $Q_2$ and $Q_3$ modes are
 respectively defined as,
 \begin{eqnarray}
 \Delta_k = \Delta_2 (- \sin \theta_k) &=& -  \sum _{k^\prime}
 \mid \tilde g_{kk^\prime} \mid
 \sin \theta_{k^\prime} ~ \sin \theta_k \nonumber \\
 & &
 \langle c_{k^\prime+Q}^{1\dagger} c_{k^\prime} ^2 +
 c_{k^\prime}^{2\dagger}
 c_{k^\prime+Q} ^1 +
 c_{k^\prime}^{1\dagger} c_{k^\prime+Q} ^2 +  c_{k^\prime+Q}^{2\dagger}
 c_{k^\prime} ^1\rangle,
 \label{eqn9a}
 \end{eqnarray}
 \begin{eqnarray}
 \Delta_k = \Delta_3 \cos  \theta_k &=& -  \sum _{k^\prime}
 \mid \tilde g_{kk^\prime} \mid
 \cos \theta_{k^\prime} ~ \cos \theta_k \nonumber \\
 & &
 \langle c_{k^\prime+Q}^{1\dagger} c_{k^\prime} ^2 +
 c_{k^\prime}^{2\dagger}
 c_{k^\prime+Q} ^1 +
 c_{k^\prime}^{1\dagger} c_{k^\prime+Q} ^2 +  c_{k^\prime+Q}^{2\dagger}
 c_{k^\prime} ^1\rangle,
 \label{eqn9b}
 \end{eqnarray}
 where $\Delta_2 (\Delta_3)$ is  the amplitude of the
 JTCCDW gap generated due to  $Q_2(Q_3)$ mode.
 Therefore, the JTCCDW Hamiltonian can be obtained from (\ref {eqntotal})
 as
 \begin{eqnarray}
 H^{Q_2(Q_3)}_{JTCCDW} & = & \sum_{k} \lambda_{1,k} (c_k^{1\dagger}c_k^1
 - c_{k+Q}^{2\dagger}c_{k+Q}^2)
 + \sum_{k} \lambda_{2,k} (c_k^{2\dagger}c_k^2
 - c_{k+Q}^{1\dagger}c_{k+Q}^1) \nonumber \\
 & + &
 \sum_k \Delta_k
 ( c_{k+Q}^{1\dagger} c_{k} ^2 + c_{k}^{2\dagger}
 c_{k+Q} ^1 +
 c_{k}^{1\dagger} c_{k+Q} ^2 +  c_{k+Q}^{2\dagger}
 c_{k} ^1),
 \end{eqnarray}
 while $\Delta_k$ is different for $Q_2$ and $Q_3$ mode according to
 equations (\ref {eqn9a}) and (\ref {eqn9b}).
 To diagonalize the Hamiltonian $ H^{Q_2( Q_3)}_{JTCCDW}$ we use 
the following two simultaneous canonical transformations :
 \begin{eqnarray}
 \alpha_{1,k} &=&\sin \frac{\phi_k}{2} ~ c^1_k + \cos \frac{\phi_k}{2}~
 c^2_{k+Q}, \nonumber \\
 \beta_{1,k} &=&
 \cos \frac{\phi_k}{2} ~ c^1_k - \sin \frac{\phi_k}{2}~ c^2_{k+Q},
 \label{eqn10}
 \end{eqnarray}
 where $\cos ( \frac{\phi_k}{2}) = \frac{1}{\sqrt 2} \left ( 1 -
 \frac{\lambda_{1,k}}
 {\sqrt{\lambda_{1,k}^2 + \Delta_k^2}}\right ) ^{\frac{1}{2}}$, 
 $~\sin ( \frac{\phi_k}{2}) = \frac{1}{\sqrt 2} \left ( 1 + \frac{\lambda_{1,k}}
 {\sqrt{\lambda_{1,k}^2 + \Delta_k^2}}\right ) ^{\frac{1}{2}}$ and
 \begin{eqnarray}
 \alpha_{2,k} &=&
 \sin \frac{\delta_k}{2} ~ c^2_k + \cos \frac{\delta_k}{2}~ c^1_{k+Q},
 \nonumber \\
 \beta_{2,k} &=&
 \cos \frac{\delta_k}{2} ~ c^2_k - \sin \frac{\delta_k}{2}~ c^1_{k+Q},
 \label{eqn11}
 \end{eqnarray}
 with $\cos ( \frac{\delta_k}{2}) = \frac{1}{\sqrt 2} \left ( 1 -
 \frac{\lambda_{2,k}}
 {\sqrt{\lambda_{2,k}^2 + \Delta_k^2}}\right ) ^{\frac{1}{2}}$ and
 $\sin ( \frac{\delta_k}{2}) = \frac{1}{\sqrt 2} 
\left ( 1 + \frac{\lambda_{2,k}}
 {\sqrt{\lambda_{2,k}^2 + \Delta_k^2}}\right ) ^{\frac{1}{2}}$.
 These transformations yield the quasi particle energies
 as $E_k^{\alpha_1}=-E_k^{\beta_1} = \sqrt {(\lambda_{1,k}^2 +
 \Delta_k^2)}$
 and $E_k^{\alpha_2}=-E_k^{\beta_2} = \sqrt {(\lambda_{2,k}^2 + \Delta_k^2)}$,
 which correspond to resulting four bands
 $\alpha_1$, $\beta_1$, $\alpha_2$ and $\beta_2$ respectively.
 The diagonalized Hamiltonian in presence of either $Q_2$ or $Q_3$ mode
 can now be written as,
 \begin{eqnarray}
 H^{Q_2(Q_3)}_{JTCCDW} =  \left [ E_k^{\alpha_1} (\alpha_{1,k}^\dagger
 \alpha_{1,k} - \beta_{1,k}^\dagger \beta_{1,k}) +
 E_k^{\alpha_2} (\alpha_{2,k}^\dagger \alpha_{2,k} -
 \beta_{2,k}^\dagger \beta_{2,k}) \right ].
 \label{eqn12}
 \end{eqnarray}
 The gap equations can be written for $Q_2$ mode as
 \begin{eqnarray}
 1 = \tilde g \sum _k {\sin^2 \theta_k} ~ \left [
 \frac{\tanh ~ (\beta~ E_k^{\alpha_1}/2)}{E_k^{\alpha_1}}
 + \frac{\tanh ~ (\beta~ E_k^{\alpha_2}/2)}{E_k^{\alpha_2}} \right ],
 \label{eqn14a}
 \end{eqnarray}
 and for the $Q_3$ mode as
 \begin{eqnarray}
 1 = \tilde g \sum _k {\cos^2 \theta_k} ~ \left [
 \frac{\tanh ~ (\beta~ E_k^{\alpha_1}/2)}{E_k^{\alpha_1}}
 + \frac{\tanh ~ (\beta~ E_k^{\alpha_2}/2)}{E_k^{\alpha_2}} \right ],
\label{eqn14b}
 \end{eqnarray}
where  for simplicity we have taken $\mid \tilde g_{kk^\prime} \mid = \tilde g$.
 It is clear that both the amplitudes
 $\Delta_2$ and $\Delta_3$ are symmetric in $k_x$ and $k_y$ sector of the
 Fermi surface, but $\Delta_k$ are highly anisotropic in $k$-space.
 The effective temperature, $\widetilde T = k_B T t$ enters in (\ref {eqn14a})
 and (\ref {eqn14b}) through $\beta = 1/{(k_B T t)}$.
 The free energy of the system can then be calculated as
 \begin{eqnarray}
 F & = & - {\frac{1}{\beta}} \sum_{k,i} \ln (1 + e ^{-\beta E^i_{k}}),
 \label{eqn15}
 \end{eqnarray}
 where $E^i_{k} = E^{\alpha_1}_k$, $E^{\beta_1}_k$,
 $E^{\alpha_2}_k$, $E^{\beta_2}_k$. Hence the heat capacity can also be
 calculated from the relation,
 \begin{eqnarray}
 C_V & = & -~{\widetilde T}~  \frac{d^2F}{d{\widetilde T}^2}.
 \label{eqn16}
 \end{eqnarray}

 We like to point out here that the four band Hamiltonian, similar to
 equation (\ref {eqn12}), is also obtained by Jackeli et al \cite{plak}
 considering only nearest neighbor Coulomb correlation in a double
 degenerate system. To explain the
nontrivial observed order in $LaMnO_3$, several studies have been reported
based on either purely Coulombic interaction or Jahn-Teller interaction
\cite{toku}.  In the framework of two orbital model, purely Coulombic
approaches produce correct spin ordered ($A$-type spin antiferromagnetic) state \cite{mae,fein}, while giving conflicting results regarding the orbital
order that co-exists with the $A$-type spin state. 
According to Hotta et al \cite{hotta}, Jahn-Teller based calculations lead to
experimentally observed $A$-type spin and $C$-type orbital order in a model
for undoped manganites. Therefore, in the present calculation focussing on 
undoped manganites, we formulate our model based on the Jahn-Teller coupling.

 The robustness of the ordered ground state is usually tested by looking at
 the response of the system to small fluctuations. In case of CDW state, it is
 well known that fluctuations of the phase and the amplitude of
 the order parameter results in the appearance of the collective modes of the
 system \cite{grun}. The amplitude mode is a measure of the CDW order parameter. In the present work, we want to calculate the spectral density function for 
the JTCCDW state, from the amplitude response function
 $\chi_{2,3}^a (q, \tilde \omega)$ for the $Q_2$ or $Q_3$ modes.
The amplitude response function is given by, 
\begin{eqnarray}
 \chi_{2,3}^a( q,\omega)&=& \sum _{k, k^\prime} f(\theta_{(k+q)},
 \theta_{(k^\prime+q)}) \nonumber \\
 & &\langle \langle~  (\sum_{i=1,2}\Psi_{ik}^\dagger (t) {\hat \tau_1}
 \Psi_{ik+q}(t)) ;
 (\sum_{i=1,2}\Psi_{ik^\prime + q }^\dagger (0) {\hat \tau_1}
 \Psi_{ik^\prime}(0))~
 \rangle \rangle ~_{\tilde \omega} ,
 \label{eqn18}
 \end{eqnarray}
 where
 \begin{eqnarray}
 f(\theta_k, \theta_{k^\prime}) & = & \sin \theta_k ~ \sin
 \theta_{k^\prime} ;
 ~~~~~for ~  Q_2- mode \nonumber \\
 & = & \cos \theta_k ~ \cos \theta_{k^\prime} ;
 ~~~~~ for~  Q_3- mode,
 \label{eqn19}
 \end{eqnarray}
 and
 $\Psi_{1,(2)~ k}^\dagger  \equiv (c^{1,(2)~\dagger }_{k}
 ~~ c^{2,(1)~\dagger}_ {k+Q} )$, are similar to
 Nambu operators. $\hat \tau_1$ is the Pauli matrix.
 Within the random phase approximation (RPA), we obtain, 
 \begin{eqnarray}
 \chi_{2,3~ RPA}^a(q,\tilde \omega)=
 \frac {\chi_{2,3}^a(q, \tilde \omega)}
 {\left [1 - \tilde g \chi_{2,3}^a (q,\tilde \omega)\right ]}.
 \label{eqn20}
 \end{eqnarray}
 The spectral density function is related to imaginary part of
 $\chi_{RPA}(\tilde \omega )$, which is given by,
 \begin{eqnarray}
 S~(0,~\widetilde \omega)~=~Im~\chi_{2,3~RPA}^a(\tilde \omega) = \frac
 {\chi_{2,3}^I (0, \tilde \omega)}
 {(1-\tilde g \chi_{2,3}^R (0, \tilde \omega))^2 + (\tilde g
 \chi_{2,3}^I(0,
 \tilde \omega))^2},
 \label{eqn21}
 \end{eqnarray}
 where $\chi_{2,3}^R (0, \tilde \omega)$ and $\chi_{2,3}^I (0,\tilde
 \omega)$
 are respectively the real and imaginary part of
 $\chi_{2,3}^a (0, \tilde \omega)$.
 By using equation (\ref{eqn18}) the expressions for 
$\chi_{2,3}^a(0, \tilde \omega)$ are obtained as,
 \begin{eqnarray}
 \chi_2^a (0,\tilde \omega) & = & 4 \sin^2 \theta_k~~ \chi (Q,\tilde
 \omega)
 \nonumber \\
 \chi_3^a (0,\tilde \omega) & = & 4 \cos^2 \theta_k~~ \chi (Q, \tilde
 \omega),
 \label{eqn22}
 \end{eqnarray}
 with
 \begin{eqnarray}
 \chi (Q, \tilde \omega) & = & \sum_k \left [ \cos^2 \phi_k ~
 \frac{E_{k}^{\alpha_1}\tanh ~ (\beta E_{k}^{\alpha_1}/2)}
 {\tilde \omega^2 ~-~ 4{E_{k}^{\alpha_1}}^2}
 + \cos^2 \delta_k ~
 \frac{E_{k}^{\alpha_2}\tanh ~ (\beta E_{k}^{\alpha_2}/2)}
 {\tilde \omega^2 ~-~ 4{E_{k}^{\alpha_2}}^2} \right ].
 \label{eqn23}
 \end{eqnarray}

 \section{Results and Discussions}
 In previous section we have formulated a theory for the charge density wave
 state in a Jahn-Teller active system. Self consistent solution of equation
 (\ref {eqn14a}) or (\ref {eqn14b}) yields $\Delta_2$ or $\Delta_3$
 respectively.
 All parameters are expressed in units of $t$ however temperature $T$ is
 expressed as $\widetilde T=k_BTt$.
 In figure 1, we show the variation of $\Delta_2$
 and $\Delta_3$ (in units of $t$) 
with temperature $\widetilde T$ in a FM spin state.
It is evident that the magnitude of JTCCDW gap decreases with increasing
temperature.
The magnitude as well as
transition temperature $\widetilde T_{CDW}$ for $Q_3$ mode is larger
than that of $Q_2$ mode for $\widetilde g~ =~ 1$. The magnitude of $\Delta_3$
is not always larger than $\Delta_2$. This is depicted in figure 2, where
variation of JTCCDW gap with $\widetilde g$ is shown for a very low
temperature. It is shown in figure 2 that $Q_2$ distortion mode opens up 
JTCCDW gap at a lower value of
$\widetilde g$  than that due to $Q_3$ mode and $\Delta_3$ is less than
$\Delta_2$ till  $\widetilde g \leq 0.6$. With increasing value of
$\widetilde g$ both
$\Delta_2$ and $\Delta_3$ increases and for values of $\widetilde g$
greater than 0.6, $\Delta_2$ is less than  $\Delta_3$.
So it is clear that the JT modes have different electron-phonon
interaction dependences. To determine the dominant JT mode
we show the variation of free energy $F$ in FM state, with $\widetilde g$
in figure 3. At very small values of $\widetilde g$, e-ph coupling 
does not alter the
energy significantly, hence presence of either of the JT modes yields the
same energy. For higher values of $\widetilde g$ the JT coupling becomes
pronounced and it is evident from figure 3  that $Q_2$ distortion mode
acquires lowest free energy for any value of $\widetilde g$.

It is well known that low temperature heat capacity ($C_V$) of CMR manganites
has a bearing on many fundamental properties like density of states, lattice
contribution and spin wave stiffness. Recent experiments on insulating
$LaMnO_3$ \cite{ghiv} show that at low temperature, $C_V/T$ vs $T^2$ plot
gives approximately straight lines. 
In an attempt to compare the lattice contributions of the specific heat in  
JTCCDW state driven by $Q_2$ or $Q_3$ mode, we show 
a plot of $C_V/{\widetilde T}$ vs $\widetilde T^2$ in figure 4 at low 
temperature. Figure 4 shows that at low temperature the $T^3$
behavior is pronounced for both the JT modes but the Debye temperatures
seem to be different.

Very recently in Raman scattering experiment on $LaMnO_3$ \cite{saitoh} by 
Saitoh et al, the orbital excitations were
observed. According to Brink \cite{brink}, the elementary excitations of
$LaMnO_3$ show both the orbital and phonon character due to
 the mixing of orbital and phonon modes. In the present system,
 JTCCDW state occurs due to softening of either $Q_2$ mode or $Q_3$ mode.
 Here charge densities at different $e_g$ orbitals depend on whether
 $Q_2$ or $Q_3$ phonon coupling is active. The $Q_3$ mode couples with the
 electron density difference of $d_{x^2-y^2}$ and $d_{3z^2-r^2}$ orbitals
 whereas the $Q_2$ mode interacts with the charge density difference between
 bonding and anti bonding orbitals formed by these two orbitals \cite{mm}.
 It is expected that the low lying collective excitations of resulting JTCCDW  
 will show different $\widetilde g$ dependence for $Q_2$ and $Q_3$ modes.
 It is evident from equation
 (\ref {eqn21}), (\ref {eqn22}) and (\ref {eqn23}) that spectral density
 function $S (0,\widetilde \omega)$ for amplitude mode at some symmetry
 points are due to any one of the JT modes.
 As example, all spectral density at symmetry
 point ($0$, $0$) is due to $Q_3$ mode while at ($\pi$, $0$)
 only $Q_2$ mode contributes to $S(0,\widetilde \omega)$. It is also
 obvious that at these symmetry points the CDW gap exists due to the JT
 mode which contributes to $S (0, \widetilde \omega)$. However,
 the $S (0, \widetilde \omega)$ as well as $\Delta_k$ at ($\pi$, $\pi/2$)
 is finite for both $Q_2$ and $Q_3$ mode. In figure 5 and figure 6 we show
 $S (0, \widetilde \omega)$ at ($\pi$, $\pi/2$) for different values of
 $\widetilde g$ in $Q_2$ and
 $Q_3$ excited CDW state respectively. With increasing $\widetilde g$ the
 peak  shifts to higher values of frequency $\widetilde \omega$
 for both the JT modes, because the amplitude of the JTCCDW gap increases
 for both the modes. It is important to note that the peak position of
 $S (0, \widetilde \omega)$ shifts more for $Q_2$ mode than that of $Q_3$
 with equal increase of $\widetilde g$. So the electron lattice
 interaction has more pronounced effect on the collective excitations
 of $Q_2$ active CDW state than that of $Q_3$ mode.

 \section{Conclusions}
 The charge density wave may have its origin due to softening of the
 breathing mode phonon or Jahn-Teller phonons. Emergence of the Jahn-Teller
 coupled charge density wave is relevant in a system where Jahn-Teller phonons
 play a crucial role to determine the orbital ordering.
 In the limit of one electron per site and strong Hund's rule
 coupling, we provide a comprehensive study of the formation of
 charge density wave emphasizing on  $e_g$ electron - Jahn-Teller coupling
 in a orbital ordered system with two kinds of
 orbitals alternating on adjacent sites in the $x-y$ plane.
 The in-plane distortion mode -- $Q_2$ is the dominant mode and it
 opens up JTCCDW gap at smaller electron-lattice interaction strength 
 than the octahedral stretching mode -- $Q_3$. The heat capacity shows $T^3$
 behavior at low temperature.  At some of the $k$-points, the effect
 of the electron-lattice interaction on collective mode such as amplitude
 mode is more pronounced when the excited Jahn-Teller mode is $Q_2$.

The present calculation is done for a system with one electron per 
lattice site and either of the JT modes is active. The divergence of the
phonon response functions corresponding to $Q_2$ or $Q_3$ mode ensures
their softening and hence the formation of CDW state. However, in presence of 
both the modes, as in real systems, it is complicated to formulate the charge
density wave instability from first principle calculation.  With hole doping 
in the present system there will be decrease of Jahn-Teller active 
$Mn^{3+}$ ions and the perfect nesting of the Fermi surface will be destroyed.
These are expected to reduce the effective electron-JT interaction as well as 
JTCCDW gap $\Delta_k$, which may yield a orbitally disordered system.
The work along these lines are in progress. \\

\noindent{\bf Acknowledgements}\\
This work is supported by SERC, DST, GOVT of India through the Fast Track 
Scheme for Young Scientists (Project Ref. No. SR/FTP/PS-29/2003). 
I would like to thank Prof. S. Yarlagadda of Saha Institute of Nuclear Physics,
Kolkata for many useful discussions.

\newpage
 {\bf Figure captions :}
\begin{itemize}
\item[\bf Fig. 1.]  Thermal variations of the $\Delta_2$ (solid line) and
 $\Delta_3$ (dashed line) in
 FM state, for $\widetilde g = 1.0$. Here, $\widetilde T = k_BTt$.

 \item[\bf Fig. 2.] Variations of $\Delta_2$ (solid line)
 and $\Delta_3$ (dashed line) with
 $\widetilde g$ at $\widetilde T =(k_BTt)= 10^{-5}$ in FM state.

 \item[\bf Fig. 3.] Variation of free energy $F$ in {\bf FM} state for
 $Q_2$ distorted CDW state (solid line) and $Q_3$ distorted CDW state
 (dashed line) with $\widetilde g$ at $\widetilde T = 0.005$.

 \item[\bf Fig. 4.]$C_V/\widetilde T$ vs $\widetilde T^2$ plot in FM state
 for $Q_2$ active (solid line) and $Q_3$ active (dashed line) CDW state
 with $\widetilde g=1.0$. Here $C_V$ is in arbitrary units and $\widetilde
 T = k_BTt$.

 \item[\bf Fig. 5.] Spectral density functions $S (0, \widetilde \omega)$,
 for
 the
 JTCCDW amplitude mode as a function of frequency $\widetilde \omega$
 (in unit of $t$) in presence of $Q_2$ distortion,
 at $k$-point ($\pi$, $\pi/2$),
 for different values of $\widetilde g=$ 0.45, 1.0, 1.5.

 \item[\bf Fig. 6.] Spectral density functions $S (0, \widetilde \omega)$,
 for the
 JTCCDW amplitude mode as a function of frequency $\widetilde \omega$
 (in unit of $t$) in presence of $Q_3$ distortion,
 at $k$-point ($\pi$, $\pi/2$),
 for different values of $\widetilde g=$ 0.45, 1.0, 1.5.

 \end{itemize}
 \noindent
 
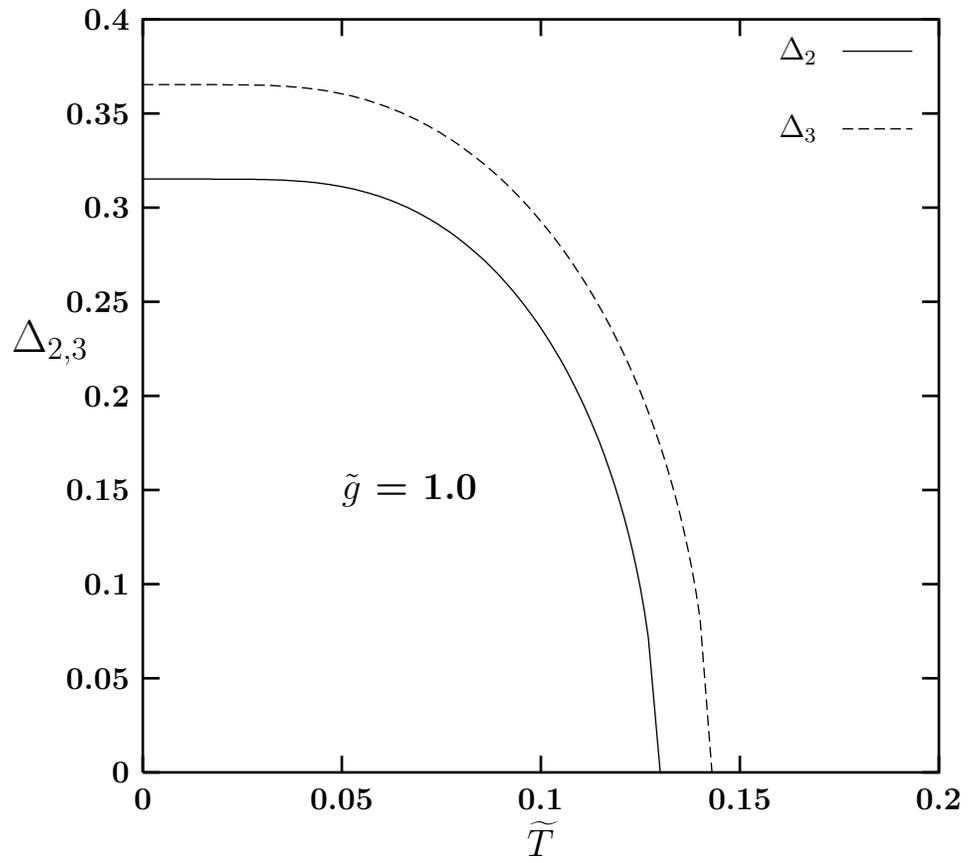
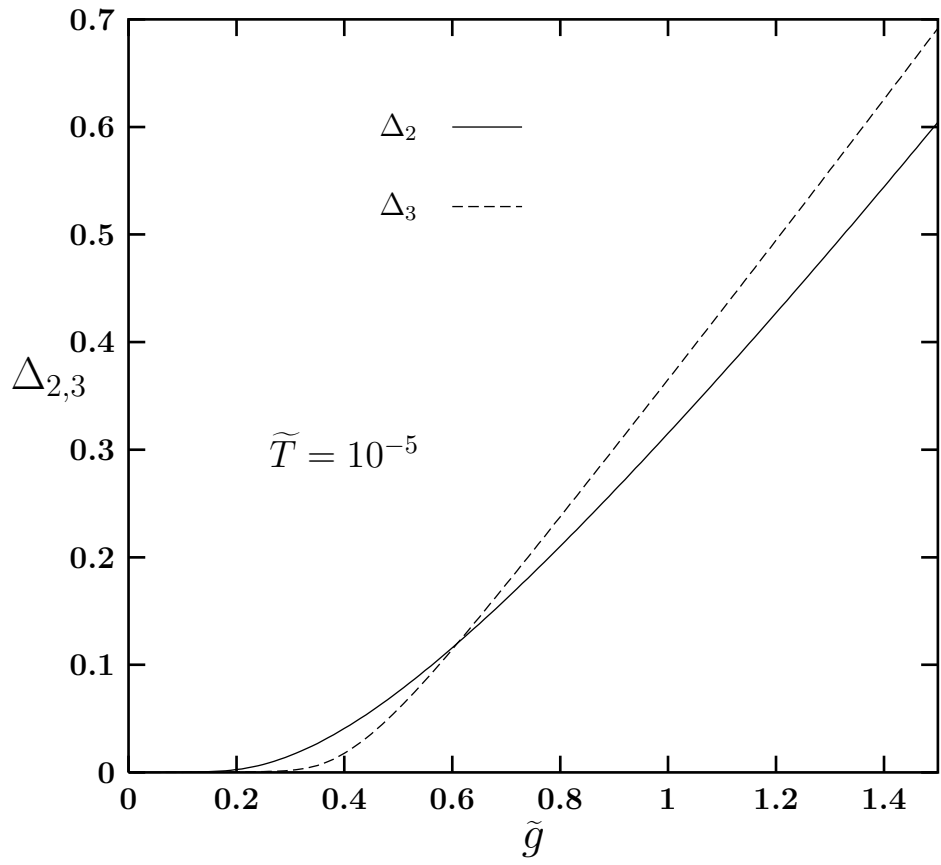
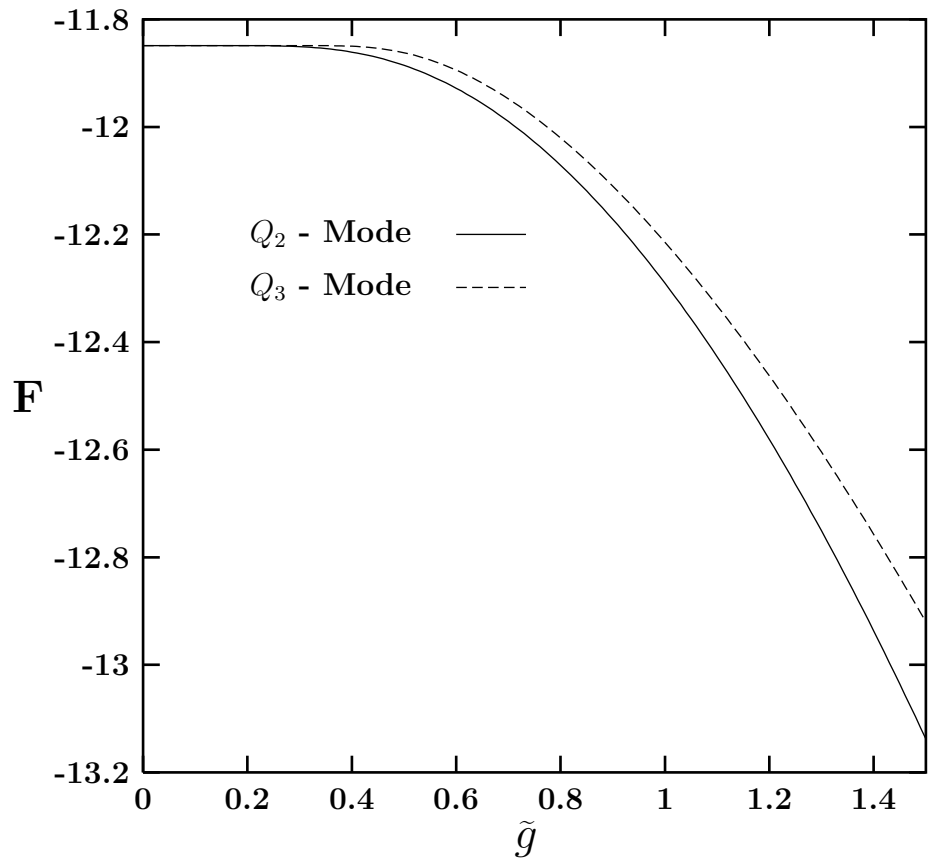
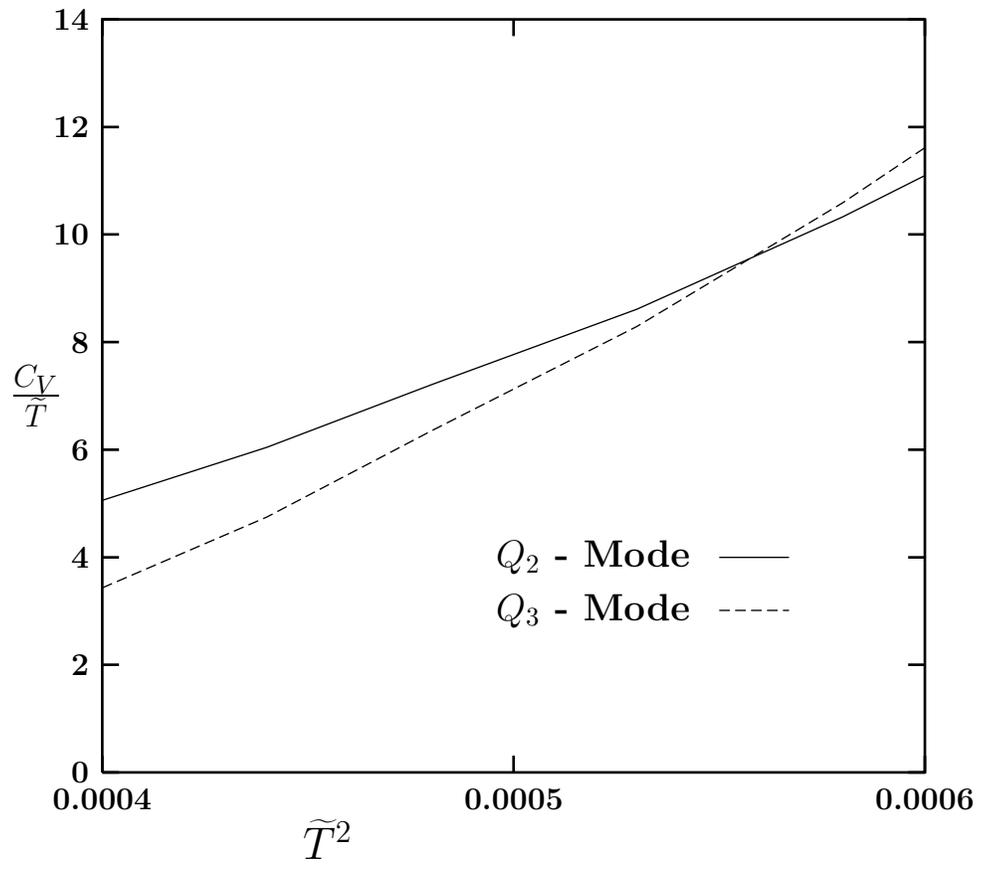
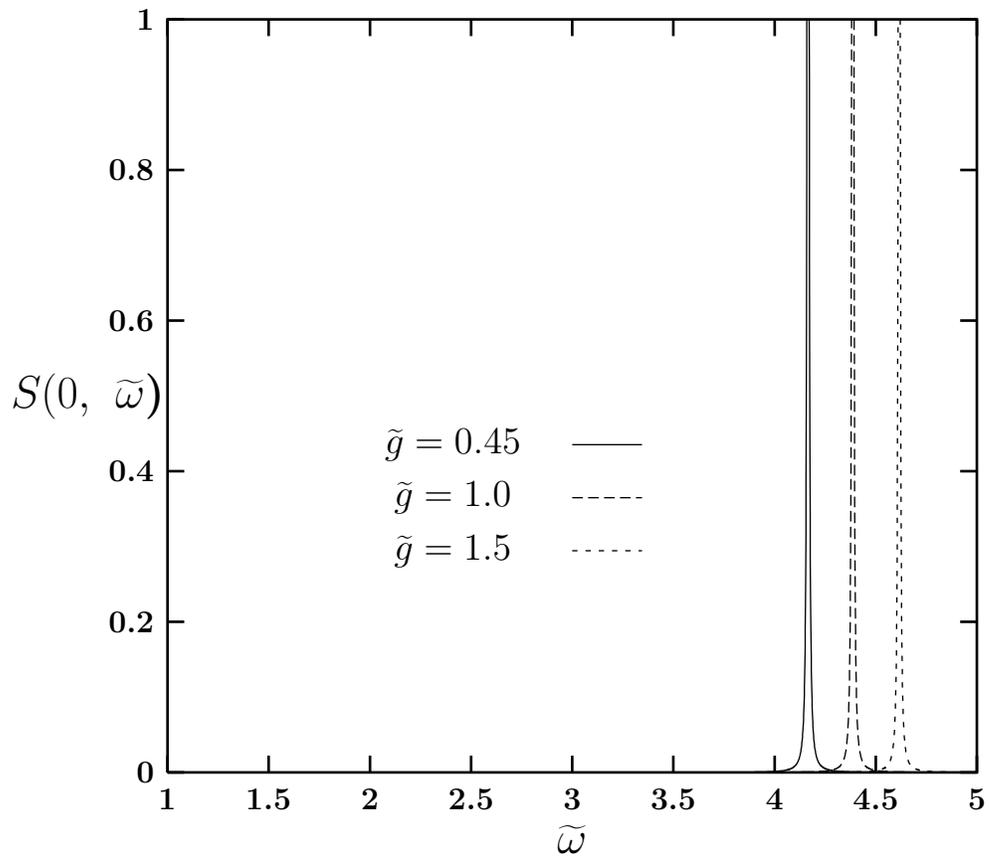
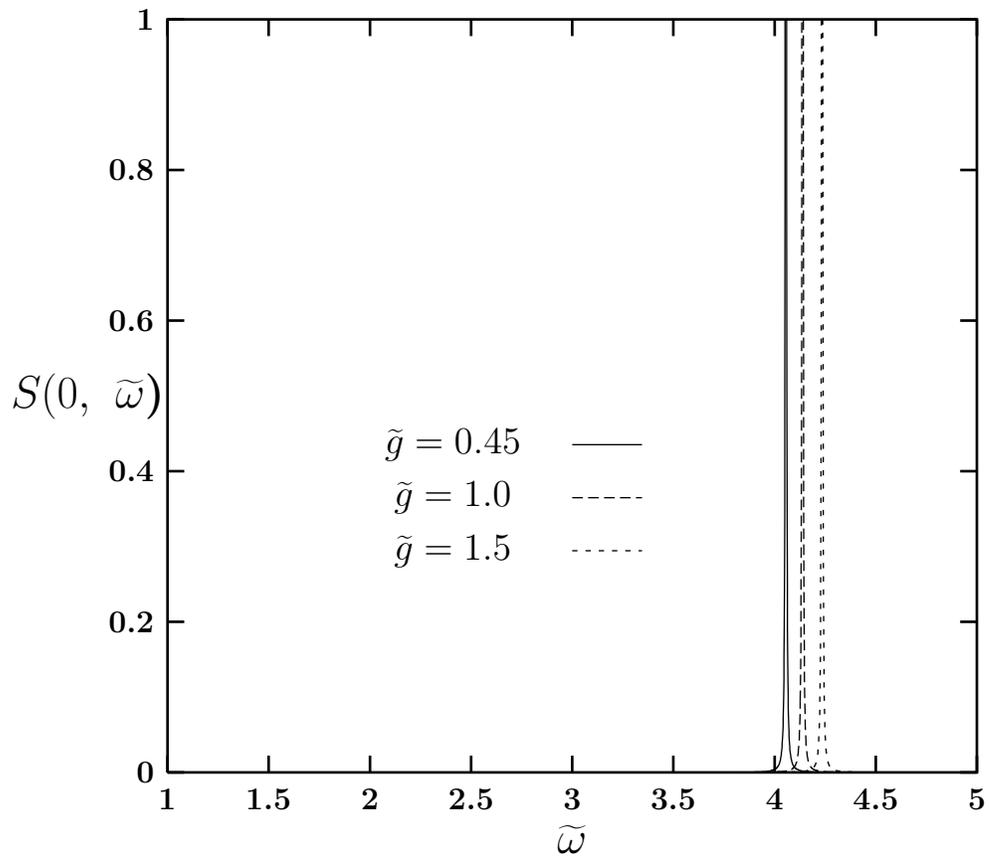
\end{document}